\documentstyle[pra,aps,amsfonts,epsfig]{revtex}
\begin{document}
\draft
\twocolumn[\hsize\textwidth\columnwidth\hsize\csname 
@twocolumnfalse\endcsname
\title{On the Biphoton Wavelength} 
\author{P. H. Souto Ribeiro}
\address{Instituto de F\'{\i}sica, 
Universidade Federal do Rio de 
Janeiro, Caixa Postal 68528, Rio de Janeiro, RJ 22945-970, Brazil} 
\date{\today}
\maketitle
\begin{abstract}
We report on an experiment showing that the wavelength of a biphoton is clearly
dependent on the measurement scheme and on the way it is defined. It is 
shown that it can take any value, depending on the control of the interferometer phase
differences. It is possible to identify the interference of the single and two-photon 
wavepackets as particular cases of the most general
interference process. The variable wavelength has no implication on the energy of the
individual photons neither on the total energy of the biphoton.
\end{abstract}
]
\section{Introduction}

The use of interferometry for measuring the wavelength of a radiation field
is probably one of its older applications. Nowadays the concept of interferometry
has been extended. It has become possible to observe experimentally the interference
for one particle and even for multiparticle wave fields. The two-photon field
produced in the parametric down-conversion, have been extensively utilized in many
of the so-called multiparticle interferometry experiments\cite{1,2}.

In this new type of interference, it is not possible to ignore the quantum aspects
of the electromagnetic field. In quantum interferometry, it is also possible to associate
interference patterns to the wavelength of a field. For single photon fields, classical and quantum interpretations of  the interference experiments lead to the same wavelength. For multi-photon 
or multi-particle fields however, the wavelength can be dependent on the way
the measurement is performed, and  a classical interpretation is no longer possible. 
Thinking of two-photon wavepackets, for example, if we can make the two photons travel 
together through
an interferometer as they were contained in one packet, we can measure a wavelength
corresponding to an entity with the energy two times larger than the single photon 
one\cite{3,4}. This concept is quite general in quantum physics and it can be extended
to any particle or field and the {\em DeBroglie} wavelength will be associated 
to the total energy of the system.

In this paper we study the two photon interference from the point of view of the
measurement of the wavelength. We present an experiment whose configuration is
capable to produce quantum interference without the use of material interferometers,
in the sense that no double-slits and no beam-splitters are used. It consists of a
transverse version of interferometers of the type of Mandel's\cite{5} and 
Zeilinger's\cite{6}. It is also similar to the interferometer presented by Klyshko et al.
in Ref.\cite{7}, but without the double-slits and with the possibility of
detecting signal and idler photons in completely independence, as it will be shown.
That is the main difference from previous transverse interferometers\cite{8,9,10,11}.
Another experiment recently performed by Fonseca et al.\cite{11b}, utilizes the same
principle for measuring a non-local wavelength for a two-photon wavepacket. The configuration
is similar to that presented by White et al.\cite{11c} for producing polarization entangled
states with high intensities, however in our case the polarization state is not entangled.
Twin photons from the parametric down-conversion process are used. 
These photons have been called {\em biphotons} as a reference to their strong 
correlation at the quantum level. The interference fringes are obtained by measuring coincidence
counts and the frequency of the oscillation of the patterns are associated to
wavelengths for the {\em biphotons}. It is shown that this frequency can be arbitrarily varied, 
depending on the
way the measurements are performed. It is also shown, that the measured wavelengths can
be assigned to single and to two-photon wavepackets, for two kinds of measurement.
A single mode quantum theory is enough to explain the behavior of the frequency
of the patterns and it is in agreement with the experimental data.

\section{Variable wavelength two-photon interference}

\begin{figure}[h]
\vspace*{5cm}
\epsfig{file=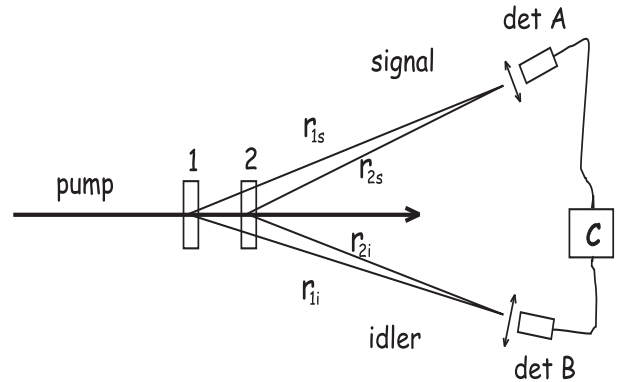,width=8cm,height=5cm}
\caption{Outline of the experiment.}
\label{fig1}
\end{figure}

Let us analyze the situation sketched in Fig. \ref{fig1}. The pump laser passes
through two nonlinear crystals, labeled crystal 1 and crystal 2. Twin photons can
be produced in each one of the crystals. Signal and idler photons produced in
crystal 1 are directed to detectors A and B respectively, so that coincidence between
signal and idler channels can be measured. Suppose that degenerate photon pairs produced
in crystal 2, can also be directed to the same detectors. This condition is simply fulfilled
by tilting crystal 2 relatively to the vertical axis. 

The situation described above is suitable for quantum interference. Note that the time of
emission of photon pairs cannot be specified, since it is a spontaneous emission and the coherence
length of the pump laser is larger than the distance between crystals. In this case, coincidence
counts produced by photon pairs originated in crystal 1 are indistinguishable from those  of 
crystal 2. Interference fringes in the coincidence counting rate can be observed, as long as
the phase difference between these two probabilities is varied.  As each crystal works like an 
extended source, no interference is observed for the individual intensities.

This interference process can be described in a simplified form with the use of a monomode
quantum approach. It is enough to explain the main properties of the coincidence patterns,
including the effective wavelength.
However, for taking into account for the degree of coherence and its consequences in the visibility 
of the fringes, a multi-mode theory would be necessary. In this work, we will restrict
ourselves to the simpler case.

The quantum state of the field produced by both crystals is given by:

\begin{equation}
\label{eq1}
|\Psi\rangle = \frac{1}{\sqrt{2}}(|1\rangle_{i1}|1\rangle_{s1}|0\rangle_{i2}|0\rangle_{s2} +
|0\rangle_{i1}|0\rangle_{s1}|1\rangle_{i2}|1\rangle_{s2}).
\end{equation}

The electric field operators for signal and idler modes at the detection planes are given by:

\begin{eqnarray}
\label{eq2}
E^{(+)}_{A} = \mbox{a}_{1s} \mbox{e}^{-i(\phi_{1s} + \mbox{k}\mbox{r}_{1s})} +
\mbox{a}_{2s} \mbox{e}^{-i(\phi_{2s} + \mbox{k}\mbox{r}_{2s})} ; \\ \nonumber
E^{(+)}_{B} = \mbox{a}_{1i} \mbox{e}^{-i(\phi_{1i} + \mbox{k}\mbox{r}_{1i})} +
\mbox{a}_{2i} \mbox{e}^{-i(\phi_{2i} + \mbox{k}\mbox{r}_{2i})},
\end{eqnarray}
where $\phi_{jx}$ is the phase of the field at the emission point, with j = 1,2 and $x$ = s,i and
k is the wave number for both signal and idler modes. r$_{jx}$ is the distance between crystal
$j$ and the detector at the $x$ side.

The coincidence counting rate can be easily calculated:

\begin{eqnarray}
\label{eq3}
C &=& | E^{(+)}_{A} E^{(+)}_{B} |\Psi\rangle |^{2} \\ \nonumber
   &=& 2[ 1 + \cos( \phi_{1i} + \phi_{1s} + \mbox{k}\mbox{r}_{1i} + \mbox{k}\mbox{r}_{1s}
\\ \nonumber
&-& \phi_{2i} - \phi_{2s} - \mbox{k}\mbox{r}_{2i} - \mbox{k}\mbox{r}_{2s})].
\end{eqnarray}

From the phase matching conditions we have that $\phi_{1i} + \phi_{1s} = \phi_{1p}$ and
$\phi_{2i} + \phi_{2s} = \phi_{2p}$, where $\phi_{1p}$ and $\phi_{2p}$ are the pump laser
phases at crystals 1 and 2 respectively. We see that one condition for observing interference
is that $\phi_{1p} - \phi_{2p}$ = const. That is to say the coherence length of the pump laser
must be larger than the distance between crystals.

Eq. \ref{eq3} can be put in the form:

\begin{equation}
\label{eq4}
C = 2 \{ 1 + \cos[ \mbox{k} (\delta_{i} + \delta_{s}) + \phi]\},
\end{equation}
where 

\begin{eqnarray}
\label{eq5}
\phi &=& \phi_{1p} - \phi_{2p} + \mbox{k} (\bar{\mbox{r}}_{1i} + \bar{\mbox{r}}_{1s} 
- \bar{\mbox{r}}_{2i} - \bar{\mbox{r}}_{2s})\, ; \\ \nonumber
\mbox{r}_{1i} &=& \bar{\mbox{r}}_{1i} + \delta_{1i}\, ; \\ \nonumber
\mbox{r}_{1s} &=& \bar{\mbox{r}}_{1s} + \delta_{1s}\, ; \\ \nonumber
\mbox{r}_{2i} &=& \bar{\mbox{r}}_{2i} + \delta_{2i}\, ; \\ \nonumber
\mbox{r}_{2s} &=& \bar{\mbox{r}}_{2s} + \delta_{2s}\, ; \\ \nonumber
\delta_{i} &=& \delta_{1i} - \delta_{2i}\, ; \\ \nonumber
\delta_{s} &=& \delta_{1s} - \delta_{2s}\, .
\end{eqnarray}

With the aid of Eqs. \ref{eq3} and \ref{eq4} it is clearly seen that the interference
fringes are sensible to phases that depend on the paths from crystals 1 and 2 to
detectors. It is worth noting that phase $\delta_{i}$ can be varied independently from 
$\delta_{s}$. Displacing signal or idler detector one can vary each one of these phases. Consider the case where $\delta_{i}$ = $\alpha \delta_{s}$.  In this case, the
variable phase  in Eq. \ref{eq4} can be written as:

\begin{equation}
\label{eq6}
C = 2 \{ 1 + \cos[ \mbox{k} (1 + \alpha) \delta_{s} + \phi]\}.
\end{equation}

The parameter $\alpha$ in the above equation can assume any value and we demonstrate
experimentally in this paper that it can be easily controlled. In fact, $\alpha$ 
is the ratio between displacements of detectors A and B. 

\section{The Experimental Set-up}

The experimental set-up shown in Fig. \ref{fig1} was implemented with a c.w. He-Cd laser operating at 442 nm. The output power was about 200 mW. It was used to pump
two 1 cm long LiIO$_{3}$ crystals. The distance between crystals was around 2 cm and
the distance between crystal 1 and both detectors A and B was nearly 1.5 m. Crystals
were cut for collinear degenerate down-conversion, that is to say the optical axis
at 37.3 degrees relatively to the input/output faces. In order to make degenerate beams
emerge from the crystal at angles different from zero, it was necessary to tilt it slightly
relatively to the vertical direction. In our case, the direction of propagation of the twin
beams were nearly 7 degrees with the pump beam direction. The output angle of the beams
produced in crystal 2 were slightly bigger, in order to achieve superposition at the
detectors with the beams originated at crystal 1.

The detectors were avalanche photodiodes inside photon counting modules (SPCM-AQ / EG\&G).
The incoming light passes through a small vertical slit (about 0.5 mm) and an AR coated lens with 25.4 mm 
focal length, before reaching the active detection area of about 0.2 mm. The modules are
mounted on X-Y translation stages, so that the transverse detection plane can be scanned with 
up to 5 $\mu$m resolution.
The output pulses were sent to counters (SR-400 and SR-620 / SRS), where single rates
were counted and the coincidence logic was implemented. Counters were controlled by a
microcomputer which was used to save data. 

\section{Experimental Results}

We have carried out coincidence interference patterns for several values of the
parameter $\alpha$ in Eq. \ref{eq6}. For doing so, we have simply changed the
relative displacement between detectors A and B.

\begin{figure}[h]
\vspace*{5cm}
\epsfig{file=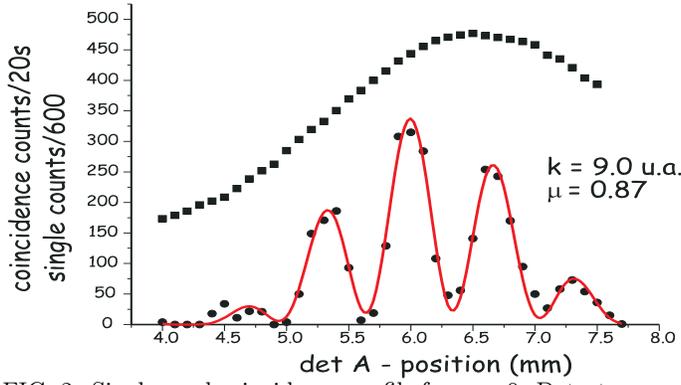,width=9cm,height=5cm}
\caption{Singles and coincidence profile for $\alpha$ = 0. Detector A is scanned
and detector B is fixed.}
\label{fig2}
\end{figure}

For $\alpha$ = 0, detector B was kept fixed, while detector A was scanned transversally 
in the horizontal plane. The single photon and the coincidence counting rates
were then registered. The results are shown in Fig. \ref{fig2}. While the singles show a
nearly gaussian profile, the coincidences show interference fringes. The visibility and
the wavevector of
the coincidence fringes were obtained by a nonlinear curve fitting with the usual
function for the double-slit interference. The wavevector for $\alpha$ = 0 will be called
k$_{0}$. It is associated to the single photon interference. Note that the absolute value
of k$_{0}$ depends on the geometry. In the analogy with a double-slit experiment, each crystal
correspond to one slit. However, the light is not emitted in the direction that would correspond 
to a central(zero order) maximum. It is emitted in a direction corresponding to a higher order.
This means that
the associated wavelength is multiplied by  a large integer, which is not important in this work. 
For this reason the wavevectors will be presented in arbitrary units. 
The procedure is repeated keeping detector
A fixed and scanning detector B. The result shows a profile similar to that of
Fig. \ref{fig2}, showing the symmetry between scans with one of the detectors fixed. 
These patterns can be interpreted as single photon wavepacket interference ones.
This is a consequence of the fact that only signal or idler paths
are changed, when only signal or idler detectors are moved individually. 

\begin{figure}[h]
\vspace*{5cm}
\epsfig{file=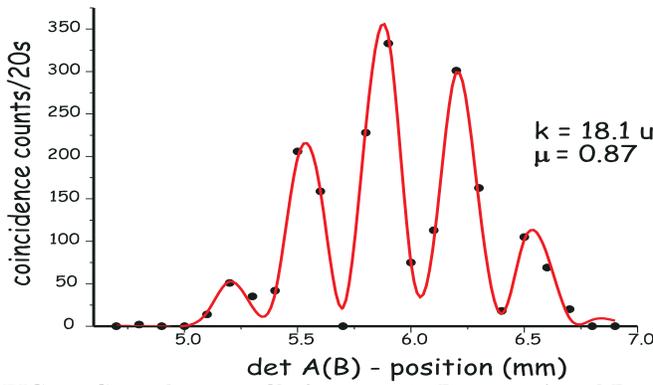,width=9cm,height=5cm}
\caption{Coincidence profile for $\alpha$ = +1. Detector A and B are scanned
simultaneously with the same velocity.}
\label{fig3}
\end{figure}

For $\alpha$ = + 1, detectors A and B were simultaneously displaced with the same velocity.
This is equivalent to saying that the detectors were displaced with equal steps. The result
is shown in Fig. \ref{fig3}. The convention used establishes that the positive $\alpha$
implies in additive phase shifts in signal and idler sides. In the experiment, this condition 
was achieved by displacing both detectors towards the pump beam. The visibility and
the wave number, that gives the frequency of the oscillations, were set as free parameters
in the fittings. As a result, we observe that the wave number for the curve in Fig. \ref{fig3}
is two times the one in Fig. \ref{fig2}. k$_{+1}$(signal side) = k$_{+1}$(idler side) = 2k$_{0}$.
This was predicted by Eq. \ref{eq6} for
$\alpha$ = + 1 and it works also for $\alpha$ = -3. The pattern of Fig. \ref{fig3}
can be interpreted as a two-photon wavepacket interference. We will address this point
again in next section.

\begin{figure}[h]
\vspace*{5cm}
\epsfig{file=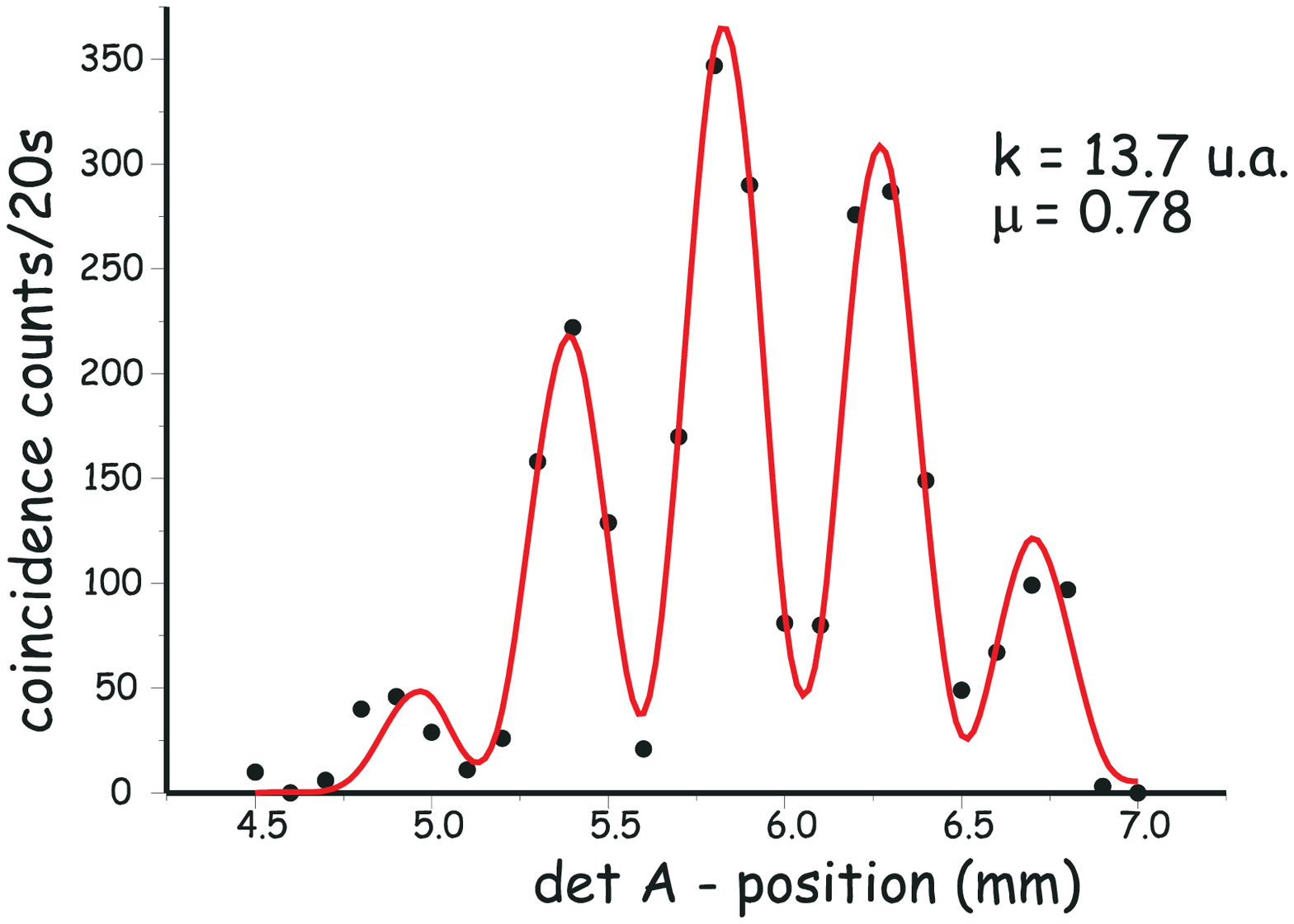,width=9cm,height=5cm}
\vspace*{5cm}
\epsfig{file=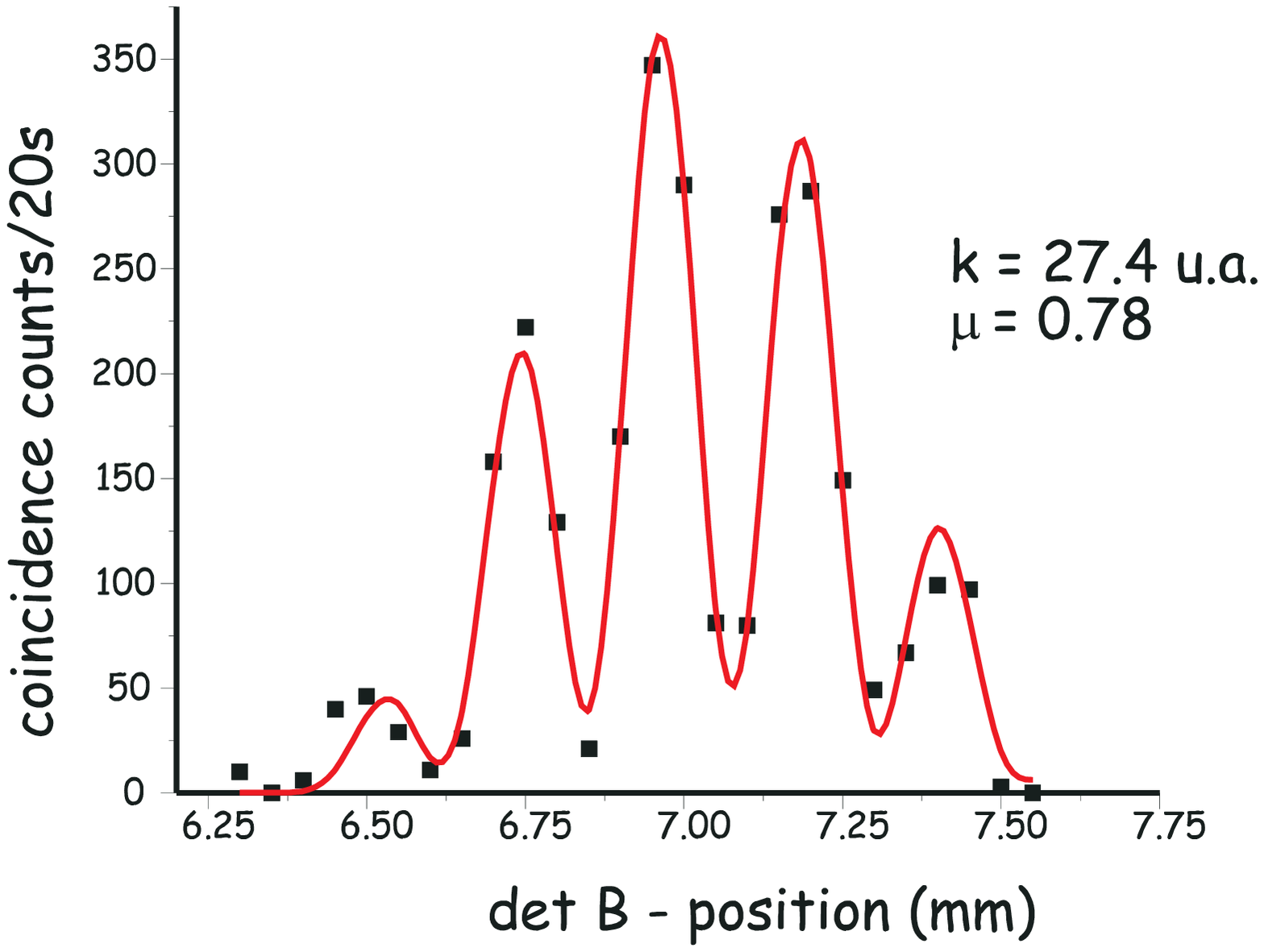,width=9cm,height=5cm}
\caption{Coincidence profiles for $\alpha$ = +$\frac{1}{2}$. Detector A and B are scanned
simultaneously with different velocities.}
\label{fig4}
\end{figure}

For $\alpha$ = $+ \frac{1}{2}$, detector A was displaced two times slower than detector B.
The displacement is performed so that detectors get together to the position corresponding
to the coincidence peak detection in previous measurements. With this procedure 
it is possible to take care for the symmetry of the interference curve. The 
results are shown in Fig. \ref{fig4}.
In Fig. \ref{fig4}a the coincidence counts are plotted as a function of the detector A position
and in Fig. \ref{fig4}b, the same coincidence counts are plotted as a function of the detector B
position. This is necessary now, because the displacements are different, we do not have a
common coordinate anymore, as in previous cases. The fitting of the curves lead to 
k$_{+\frac{1}{2}}$(signal side) = $\frac{3}{2}$  k$_{0}$, which corresponds to $\alpha$ =  + $\frac{1}{2}$  and k$_{+\frac{1}{2}}$(idler side) = 3 k$_{0}$ which corresponds to 
$\alpha_{i}$ = + 2 when the phase shift is written in terms
of the idler coordinates. This is in agreement with Eq. \ref{eq6}.

\begin{figure}[h]
\vspace*{5cm}
\epsfig{file=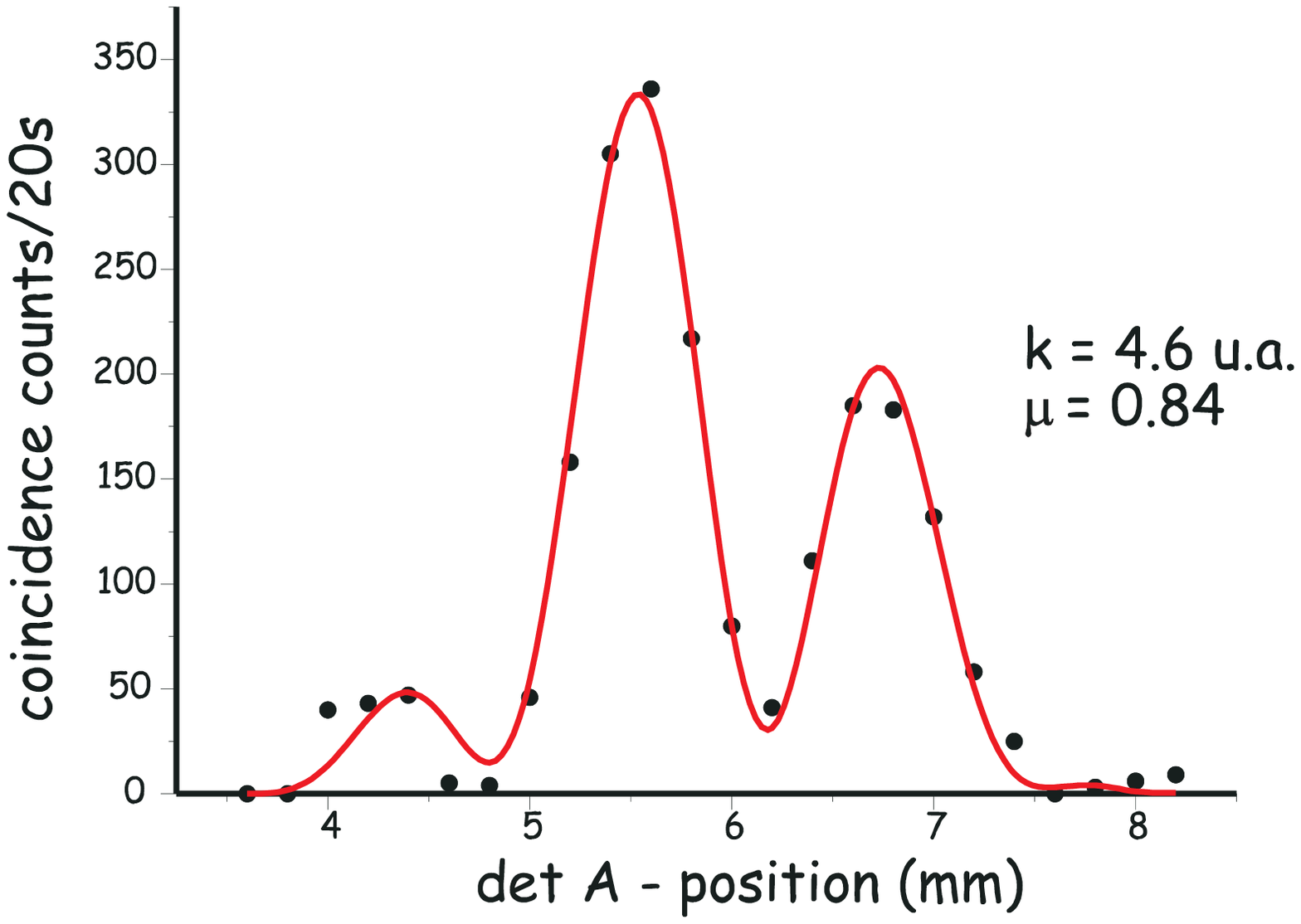,width=9cm,height=5cm}
\vspace*{5cm}
\epsfig{file=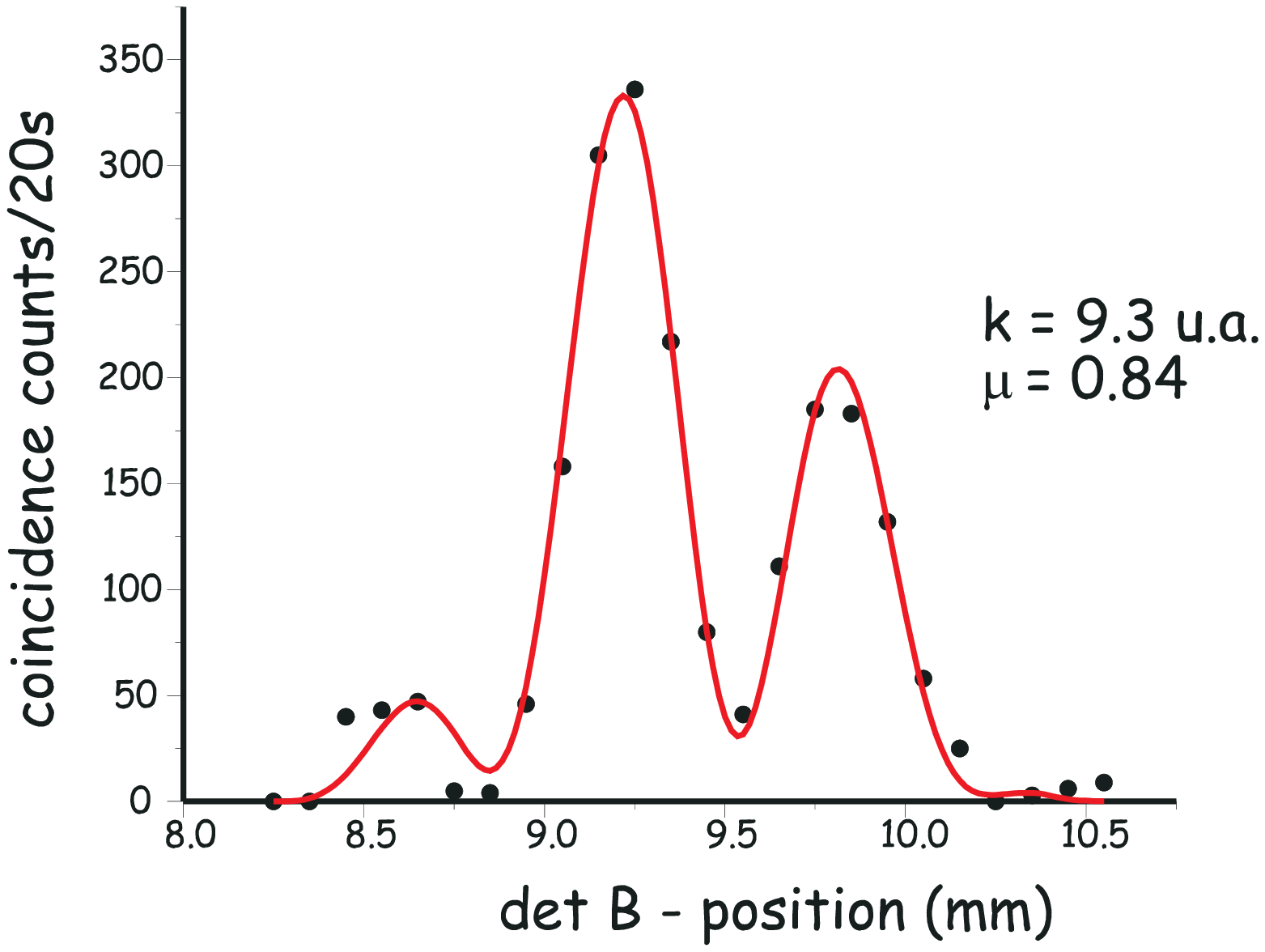,width=9cm,height=5cm}
\caption{Coincidence profiles for $\alpha$ = +$\frac{1}{2}$. Detector A and B are scanned
simultaneously with different velocities.}
\label{fig5}
\end{figure}

For $\alpha$ = - $\frac{1}{2}$, detector A is still displaced two times slower than detector B,
but now the negative sign indicates that the sense of one of the displacements is changed.
Detector A moves towards the pump beam while detector B move backwards the pump beam.
The results are shown in Fig, \ref{fig5}. In Fig. \ref{fig5}a the coincidence counts are plotted
as a function of the detector A position while in Fig. \ref{fig5}b they are plotted as a function
of the  detector B position. From the point of view of detector A $\alpha$ = - $\frac{1}{2}$
and the wavevector is k$_{-\frac{1}{2}}$(idler side) = $\frac{1}{2}$ $k_{0}$. 
From the point of view of detector B
$\alpha_{s}$ = - 2 and k$_{-\frac{1}{2}}$(signal side) = $k_{0}$.

\section{Discussion}

From Eq. \ref{eq6} and experimental results presented in previous section, it is
clear that the wavelength of the biphoton can be continuously varied and that it can
in principle assume any value, even a fraction or a multiple of the single photon one. 
This fact can be explained
by the phase entanglement between signal and idler photons. It also seems that 
it has no implications on the energy of the individual photons neither on the total energy
of the biphotons. It is clear from all experiments utilizing twin photons from the
parametric down-conversion, that the process behind all quantum effects is {\em entanglement}.
The entanglement is a consequence of a process that we could call {\em transfer of spectrum}
from the pump beam to the biphotons. A particular case of that is the {\em transfer
of angular spectrum} described in Ref. \cite{11} dealing with the transverse
degrees of freedom of the field. In the experiment we present here, it is nice to be
able to understand the main features in connection with entanglement and transfer of
angular spectrum for a simple case utilizing a monomode approach. This is one of the
virtues of the interferometer presented.

It is interesting however, to analyze some possible interpretations. When $\alpha$ = 0,
one detector is fixed and the other one is moved. In this case, the experiment can be 
viewed as a single photon wavepacket interference because the phase difference depends
only on trajectories for the same photon. When $\alpha$ = + 1 or
$\alpha$ = -2, both detectors are moved simultaneously with the same velocity. 
In this case, the experiment
corresponds to an unfolded version of a two-photon wavepacket interference. Note that
in one of the previous experiments\cite{4}, it was necessary to prepare the state of the field
by manipulating the pump beam, in order to avoid single photon interference. In the
present configuration, each crystal plays the role of one slit in the analogy with a
double slit experiment. However, each twin photon pair is {\em always}
(the probability is much bigger)  emitted by the same crystal
and {\em never} (the probability is much smaller)  by different ones. 
This corresponds to having both  photons passing through
one of the slits and never one photon through each slit. For this reason, it is not necessary
to change the pump laser beam profile. But the analogy is complete.

When $\alpha$ = + $\frac{1}{2}$, for example, it is not possible to assign some physical
meaning to the wavelength observed by one of the detectors anymore.
For the same set of measurements, the wavevector is three times larger than the single photon one (corresponding to a wavelength three times smaller than the single photon one) from the point
of view of  the conjugated detector. This result shows that assigning a wavelength to
{\em something} we call biphoton can be dangerous.
The energy of each individual photon
is not changed during the interference process, neither during the detection process, and the
discussion about this  {\em apparent} wavelength may turn into speculation. However,
diffraction properties are actually changed and that may have consequences
in imaging and other applications.

\section{Conclusion}

A two-photon interferometer without double-slits and without beam-splitters is
presented. The frequency of the oscillations of the coincidence interference patterns
is varied. These frequencies are associated to the single and to the two-photon
wavepackets. Frequencies that are fractions and multiples of the single photon one
are observed, in agreement with theory.
These variable frequency oscillations are well understood in terms of the quantum
theory and two particle entanglement. Some applications for that can be envisaged for
example in the new field of {\em quantum imaging}.

\section{ACKNOWLEDGMENT}
We thank Drs. Carlos Monken and Sebasti\~ao de P\'adua for lending one
of the detectors and also for many helpfull discussions.
Financial support was provided by Brazilian 
agencies CNPq, PRONEX, FAPERJ and FUJB. 

\end{document}